\documentclass[12pt]{article}
\usepackage{a4wide,epsfig,psfrag,amsmath,amssymb,cite,scalefnt}
\usepackage{color}
\usepackage{subcaption} 

\graphicspath{ {figs/} }

\def\({\left(}
\def\){\right)}
\def\slnA{\ln2}
\def\slnB{\ln^22}
\def\slnC{\ln^32}
\def\slnD{\ln^42}
\def\slnE{\ln^52}

\def\zB{\zeta(2)}
\def\zC{\zeta(3)}
\def\zD{\zeta(4)}
\def\zE{\zeta(5)}
\def\zF{\zeta(6)}

\def\lmm{\ln\frac{\mu^2}{m_h^2}}
\def\lmmB{\ln^2\frac{\mu^2}{m_h^2}}
\def\lmmC{\ln^3\frac{\mu^2}{m_h^2}}
\def\lmmD{\ln^4\frac{\mu^2}{m_h^2}}
\def\lmmOS{\ln\frac{\mu^2}{M_h^2}}
\def\lmmOSB{\ln^2\frac{\mu^2}{M_h^2}}
\def\lmmOSC{\ln^3\frac{\mu^2}{M_h^2}}
\def\lmmOSD{\ln^4\frac{\mu^2}{M_h^2}}
\def\api{\left(\frac{\alpha_s^{(n_f)}}{\pi}\right)}
\def\apinl{\left(\frac{\alpha_s^{(n_l)}}{\pi}\right)}

\allowdisplaybreaks

\newcommand{\gsim}{\;\rlap{\lower 3.5 pt \hbox{$\mathchar \sim$}} \raise 1pt
 \hbox {$>$}\;}
\newcommand{\lsim}{\;\rlap{\lower 3.5 pt \hbox{$\mathchar \sim$}} \raise 1pt
 \hbox {$<$}\;}

\begin{document}

\title{\vskip-3cm{\baselineskip14pt
    \begin{flushleft}
      \normalsize TTP15-05
  \end{flushleft}}
  \vskip1.5cm
  Decoupling of heavy quarks at four loops and effective Higgs-fermion
  coupling
}

\author{
  Tao Liu and Matthias Steinhauser
  \\[1em]
  {\small\it Institut f{\"u}r Theoretische Teilchenphysik}\\
  {\small\it Karlsruhe Institute of Technology (KIT)}\\
  {\small\it 76128 Karlsruhe, Germany}  
}
  
\date{}

\maketitle

\thispagestyle{empty}

\begin{abstract}

  We compute the decoupling constant $\zeta_m$ relating light quark masses of
  effective $n_l$-flavour QCD to $(n_l+1)$-flavour QCD to four-loop order.
  Immediate applications are the evaluation of the $\overline{\rm MS}$ charm
  quark mass with five active flavours and the bottom quark mass
  at the scale of the top quark or even at GUT scales.
  With the help of a low-energy theorem $\zeta_m$ can be used to obtain
  the effective coupling of a Higgs boson to light quarks with five-loop
  accuracy. We briefly discuss the influence on $\Gamma(H\to b\bar{b})$.
  
  \medskip


\end{abstract}

\thispagestyle{empty}




\section{Introduction and notation}

Perturbative calculations in QCD are quite advanced and have reached, at least
for some observables, the four and even five-loop level (see
Refs.~\cite{Baikov:2015tea,Chetyrkin:2015mxa} for a recent review). This
concerns in particular the renormalization group functions which have been
computed at four loops in
Refs.~\cite{Chetyrkin:1997dh,Vermaseren:1997fq,vanRitbergen:1997va,Czakon:2004bu,Chetyrkin:2004mf}.
The first five-loop result has been obtained recently in
Ref.~\cite{Baikov:2014qja} where the quark mass anomalous dimension has been
computed to this order.

In order to consistently relate the quark masses and strong coupling
constant evaluated at different energy scales, both the
renormalization group functions and also the decoupling relations
have to be available. The latter take care of integrating out heavy quark fields. 
In fact, $N$-loop running goes along with $(N-1)$-loop decoupling.
Thus, besides the five-loop anomalous dimensions also the
four-loop decoupling relations are needed. In
Refs.~\cite{Schroder:2005hy,Chetyrkin:2005ia} a first step has been
undertaken in this direction and the four-loop decoupling constant for
$\alpha_s$ has been computed (although the five-loop beta function is
not yet available). In this paper we complement the result by
computing the four-loop corrections to the decoupling constant for the
light quark masses, which supplements the five-loop result for
$\gamma_m$~\cite{Baikov:2014qja}.

In Ref.~\cite{Chetyrkin:1997un} a formalism has been derived which
allows for an effective calculation of the $N$-loop decoupling constants
with the help of $N$-loop vacuum integrals. In the following we 
present the formulae which are relevant for the calculation of the
quark mass decoupling constant.

The bare decoupling constant $\zeta_m^0$ is defined via the relation
\begin{eqnarray}
  m_q^{0\prime}&=&\zeta_m^0m_q^0\,,
  \label{eq::dec}
\end{eqnarray}  
where $m_q^0$ and $m_q^{0\prime}$ are the bare quark mass parameters in the
full $n_f$- and effective $n_l(\equiv n_f-1)$-flavour theory.
Introducing the renormalization constants in both theories
leads to the equation
\begin{eqnarray}
  m_q^\prime(\mu)&=&
  \frac{Z_m}{Z_m^\prime}\zeta_m^0m_q(\mu) =\zeta_m m_q(\mu)\,,
  \label{eq::decm}
\end{eqnarray}
which relates finite quantities and defines $\zeta_m$.
Note that primed quantities depend on $\alpha_s^{(n_l)}$ and
non-primed quantities on $\alpha_s^{(n_f)}$.
Four-loop results for $Z_m$ and $Z_m^\prime$
can be found in
Refs.~\cite{Chetyrkin:1997dh,Vermaseren:1997fq,Chetyrkin:2004mf}
and $\zeta_m^0$ can be computed with the help of
\begin{eqnarray}
  \zeta_m^0&=&\frac{1-\Sigma_S^{0h}(0)}{1+\Sigma_V^{0h}(0)}\,,
  \label{eq::zeta0}
\end{eqnarray}
where $\Sigma_S^{0h}(0)$ and $\Sigma_V^{0h}(0)$ are the
scalar and vector parts of the light-quark self energy evaluated
at zero external momentum. The superscript ``h'' reminds that one has to
consider only the hard part which involves at least one propagator of the heavy
quark.

In the next Section we discuss the calculation of $\zeta_m^0$ and its
renormalization to arrive at $\zeta_m$. Section~\ref{sec::let} applies a
low-energy theorem to derive, from the four-loop result of $\zeta_m$, the
effective Higgs-fermion coupling constant to five-loop order. We summarize our
findings in Section~\ref{sec::con}.


\section{Decoupling for light quark masses}

\begin{figure}[t]
  \centering
  \includegraphics[bb = 100 340 500 510, width=.9\textwidth]{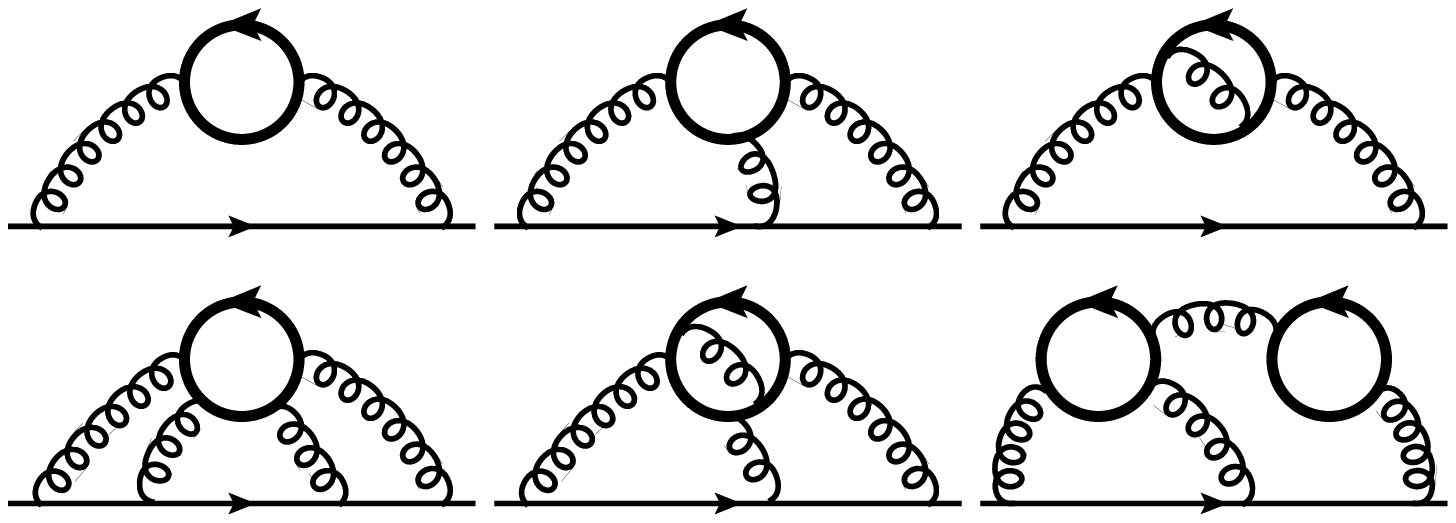}
  \caption[]{\label{fig::diags_zetam}Sample Feynman diagrams
    contributing to the hard part of the light-quark 
    propagator up to four loops. Solid and curly lines denote quarks and
    gluons, respectively. At least one of the closed fermion loops
    needs to be the heavy quark.}
\end{figure}

In this section, we compute the decoupling constant $\zeta_m^0$ and combine it
with the four-loop result for $Z_m$ to obtain the finite quantity $\zeta_m$.
The computation of $\zeta_m^0$ requires the knowledge of the hard contribution
to the scalar and vector part of the light-quark propagator, see
Fig.~\ref{fig::diags_zetam} for sample Feynman diagrams.
The first non-vanishing contribution arises at two loops where one diagram
contributes. At three-loop order there are 25 and at four loops we have 765
Feynman diagrams.

The perturbative expansion of Eq.~(\ref{eq::zeta0}) to four loops
leads to
\begin{eqnarray}
  \zeta_m^0 &=& 1 - \Sigma_S^{0h}(0)- \Sigma_V^{0h}(0) 
  + \Sigma_V^{0h}(0) \left[ \Sigma_S^{0h}(0) + \Sigma_V^{0h}(0) \right]
  + \ldots
  \,,
  \label{eq::zetam0}
\end{eqnarray}
where in the last term on the right-hand side only two-loop expressions 
for $\Sigma_S^{0h}(0)$ and $\Sigma_V^{0h}(0)$ have to be inserted.

We generate the Feynman diagrams with the help of {\tt
  QGRAF}~\cite{Nogueira:1991ex}. {\tt
  FORM}~\cite{Vermaseren:2000nd,Kuipers:2012rf} code is then generated by
passing the output via {\tt q2e}~\cite{Harlander:1997zb,Seidensticker:1999bb},
which transforms Feynman diagrams into Feynman amplitudes, to {\tt
  exp}~\cite{Harlander:1997zb,Seidensticker:1999bb}.
After processing the
latter one obtains the result as a linear combination of scalar functions
which have a one-to-one relation to the underlying topology of the diagram.
The functions contain the 
exponents of the involved propagators as arguments. At this point one has a
large number of different functions. Thus, in the next step one passes them to a
program which implements the Laporta algorithm~\cite{Laporta:2001dd} and
performs a reduction to a small number of so-called master integrals.  We use,
for the latter step, the {\tt C++} program {\tt FIRE}~\cite{Smirnov:2014hma}.
Our four-loop result is expressed in terms of 13 master integrals
which we take from Ref.~\cite{Lee:2010hs} (see
also~\cite{Laporta:2002pg,Schroder:2005va,Chetyrkin:2006dh} and references
therein). All $\epsilon$ coefficients are known analytically in the
literature except the $\epsilon^3$ term of integral $J_{6,2}$ 
(in the notation from Ref.~\cite{Lee:2010hs}) which has been
provided from~\cite{priv_R.Lee}.

Note that for our calculation we have used a general gauge parameter $\xi$ of
the gluon propagator. At four loops, in intermediate steps terms
up to order $\xi^6$ are present, however, in the final result for $\zeta_m^0$
all $\xi$ terms drop out. The last term on the right-hand side of
Eq.~(\ref{eq::zetam0}) is separately $\xi$-independent since at two loops
$\Sigma_S^{0h}(0)$ and $\Sigma_V^{0h}(0)$ are individually $\xi$-independent.
The results up to three-loop order have been checked with the
help of {\tt MATAD}~\cite{Steinhauser:2000ry} which avoids the use of 
the program {\tt FIRE} since it implements the explicit solution of the
recurrence relations.

To obtain $\zeta_m^0$ we have to renormalize $\alpha_s$ and the heavy quark
mass $m_h$ to two-loop order. The corresponding $\overline{\rm MS}$
counterterms are well-known (see, e.g.. Ref.~\cite{Chetyrkin:2004mf}).
$\zeta_m^0$ still contains poles in $\epsilon$ which are removed by
multiplying with the factor $Z_m/Z_m^\prime$ (see, Eq.~(\ref{eq::decm})) which
is needed to four-loop
order~\cite{Chetyrkin:1997dh,Vermaseren:1997fq,Chetyrkin:2004mf}.  Note that
$Z_m^\prime$ depends on the strong coupling constant of the effective theory,
$\alpha_s^{(n_l)}$, whereas $Z_m$ and $\zeta_m^0$ are expressed in terms of
$\alpha_s^{(n_l+1)}$.  In order to achieve the cancellation of the $\epsilon$
poles the same coupling constant has to be used in all three quantities.  We
have decided to replace $\alpha_s^{(n_l)}$ in favour of $\alpha_s^{(n_l+1)}$
which is done using the corresponding decoupling constant $\zeta_{\alpha_s}$
up three-loop order~\cite{Chetyrkin:1997un}. Note, however, that higher
order terms in $\epsilon$ are also needed since $\zeta_{\alpha_s}$ gets multiplied
by poles present in $Z_m^\prime$. Up to two-loop order they can be found in
Refs.~\cite{Grozin:2007fh,Grozin:2011nk}; the three-loop terms of order
$\epsilon$ can be extracted from
Refs.~\cite{Schroder:2005hy,Chetyrkin:2005ia}.

Our final result for the decoupling constant parametrized in terms of the
$\overline{\rm MS}$ heavy quark mass, $m_h\equiv m_h(\mu)$, reads
\begin{eqnarray}
  \lefteqn{\zeta_m^{\overline{\rm MS}} =}
  \nonumber\\&& \!\!\!\!\!\!\!\!\!\!\!\!
 1 + \api^2 \(
           \frac{89}{432}
          - \frac{5}{36}\lmm
          + \frac{1}{12}\lmmB\)   
          + \api^3 \left[
           \frac{2951}{2916}
          + \frac{1}{9}\zB\slnB \right.\nonumber\\&&{}  
          - \frac{1}{54} \slnD  
          - \frac{407}{864}\zC
          + \frac{103}{72} \zD
          - \frac{4}{9} a_4
          -\(\frac{311}{2592}
             + \frac{5}{6}\zC \)\lmm 
          + \frac{175}{432}\lmmB \nonumber\\&&{} \left.  
          + \frac{29}{216}\lmmC  
          + n_l\(
           \frac{1327}{11664}
          - \frac{2}{27}\zC
          - \frac{53}{432}\lmm
          - \frac{1}{108}\lmmC \) \right] \nonumber\\&&{}
          + \api^4 \Bigg[
             \frac{131968227029}{3292047360}
           - \frac{1924649}{4354560}\slnD
           + \frac{59}{1620}\slnE
           + \frac{1924649}{725760}\zB\slnB \nonumber\\&&{}       
           - \frac{59}{162}\zB\slnC       
           - \frac{353193131}{40642560}\zC
           + \frac{1061}{576}\zC^2
           + \frac{16187201}{580608}\zD
           - \frac{725}{108}\zD\slnA
           \nonumber\\&&{}
           - \frac{59015}{1728}\zE  
           - \frac{3935}{432}\zF     
           - \frac{1924649}{181440}a_4 
           - \frac{118}{27}a_5
          +\(- \frac{2810855}{373248}
          - \frac{31}{216}\slnD
\right.\nonumber\\&&{} \left. 
          + \frac{31}{36}\zB\slnB  
          - \frac{373261}{27648}\zC 
          + \frac{4123}{288}\zD  
          + \frac{575}{72}\zE
          - \frac{31}{9}a_4 \)\lmm  
\nonumber\\&&{}
          +\( \frac{51163}{10368}
          - \frac{155}{48}\zC \)\lmmB 
          + \frac{301}{324}\lmmC 
          + \frac{305}{1152}\lmmD   
\nonumber\\&&{}
          +n_l\(
          - \frac{2261435}{746496}
          + \frac{49}{2592}\slnD   
          - \frac{1}{270}\slnE      
          - \frac{49}{432}\zB\slnB
          + \frac{1}{27}\zB\slnC    
\right.\nonumber\\&&{}\left.
          - \frac{1075}{1728}\zC
          - \frac{1225}{3456}\zD  
          + \frac{49}{72}\zD\slnA
          + \frac{497}{288}\zE     
          + \frac{49}{108}a_4   
          + \frac{4}{9}a_5         
\right.\nonumber\\&&{}\left.
          + \(\frac{16669}{31104}
          + \frac{1}{108}\slnD
          - \frac{1}{18}\zB\slnB
          + \frac{221}{576}\zC     
          - \frac{163}{144}\zD   
          + \frac{2}{9}a_4     \)\lmm  \right. \nonumber\\&&{} \left.
          - \frac{7825}{10368}\lmmB 
          - \frac{23}{288}\lmmC
          - \frac{5}{144}\lmmD \)  
          +n_l^2\(
           \frac{17671}{124416}  
          - \frac{5}{864}\zC
        \right. \nonumber\\&&{} \left.
          - \frac{7}{96}\zD 
          +\( -\frac{3401}{46656}
          + \frac{7}{108}\zC \)\lmm
          +\frac{31}{1296}\lmmB
          +\frac{1}{864}\lmmD \) 
          \Bigg]
\nonumber\\
&\stackrel{\mu=m_h}{=}&\mbox{}
1
+ \left(\frac{\alpha_s^{(n_f)}(m_h)}{\pi}\right)^2 0.2060
+ \left(\frac{\alpha_s^{(n_f)}(m_h)}{\pi}\right)^3
\left(1.848 + 0.02473 n_l\right)
\nonumber\\&&\mbox{}
+ \left(\frac{\alpha_s^{(n_f)}(m_h)}{\pi}\right)^4
\left(6.850 - 1.466 n_l + 0.05616 n_l^2\right)
\,,
\label{eq::zetamMS}
\end{eqnarray}
with $\alpha_s^{(n_f)}\equiv \alpha_s^{(n_f)}(\mu)$.  In the analytic
expression $\zeta(n)$ denotes the Riemann zeta function evaluated at $n$
and $a_n=\mbox{Li}_n(1/2)$.

Often it is convenient to express $\zeta_m$ in terms of the 
on-shell heavy quark mass, $M_h$. The corresponding analytic expressions are
obtained from Eq.~(\ref{eq::zetamMS}) with the help of the 
two-loop relation between $m_h(\mu)$ and $M_h$ which can be found in
Refs.~\cite{Gray:1990yh,Chetyrkin:1999qi,Melnikov:2000qh}.
We refrain from
showing the corresponding analytic result and restrict the presentation to
the numerical expression which is given by
\begin{eqnarray}
  \zeta_m^{\rm OS} &=& 
  1 + \api^2 \left(0.2060 - 0.1389\lmmOS +  0.08333\lmmOSB\right)  
  \nonumber\\&&\mbox{}
  +\api^3 \left[ 1.477- 0.9550\lmmOS + 
    0.7384\lmmOSB + 0.1343\lmmOSC 
  \nonumber\right.\\&&\left.\mbox{}
    + 
    n_l\left(0.02473 - 0.1227\lmmOS - 
    0.009259\lmmOSC\right) \right] 
  + \api^4  \bigg[\left. 0.2233
    \nonumber\right.\\&&\left.\mbox{}
    + 2.674\lmmOS + 
    6.227\lmmOSB + 2.165\lmmOSC + 
    0.2648\lmmOSD 
    \nonumber\right.\\&&\left.\mbox{}
    + n_l \left(-1.504- 0.6470\lmmOS - 
    0.9260\lmmOSB - 0.1632\lmmOSC 
    \nonumber\right.\right.\\&&\left.\left.\mbox{}
    - 
    0.03472\lmmOSD\right) +
    n_l^2 \left(0.05616 + 0.005016\lmmOS + 
    0.02392\lmmOSB 
    \nonumber\right.\right.\\&&\left.\left.\mbox{}
    + 0.001157\lmmOSD\right)
  \bigg]\right.
  \,.
\end{eqnarray}
On the webpage~\cite{progdata} we provide analytic results in
computer-readable form for a general $SU(N_c)$ gauge group.

In the remaining part of this section we discuss two applications which
involve the evaluation of light quark masses at high scales. In the first one
we compute the running bottom quark mass at the scale $\mu=M_t$, where
  $M_t$ is the top quark pole mass. $m_b(M_t)$ appears
as an intermediate step in analyses concerned with Yukawa coupling
unification.  Here the role of the heavy quark is taken over by the top quark.
In the second application we cross the bottom threshold and evaluate the charm
quark mass for $\mu=M_Z$ using $m_c^{(4)}(3~\mbox{GeV})$ as input.  As input
parameters for the numerical analyses we
use~\cite{Agashe:2014kda,Chetyrkin:2009fv}
\begin{eqnarray}
  \alpha_s^{(5)}(M_Z) &=& 0.1185\,,\nonumber\\
  m_b^{(5)}(m_b^{(5)}) &=& 4.163~\mbox{GeV}\,,\nonumber\\
  m_c^{(4)}(3~\mbox{GeV}) &=& 0.986~\mbox{GeV}\,.
\end{eqnarray}

\begin{figure}[t]
  \centering
  \begin{subfigure}[b]{.7\textwidth}
    \includegraphics[width=\textwidth]{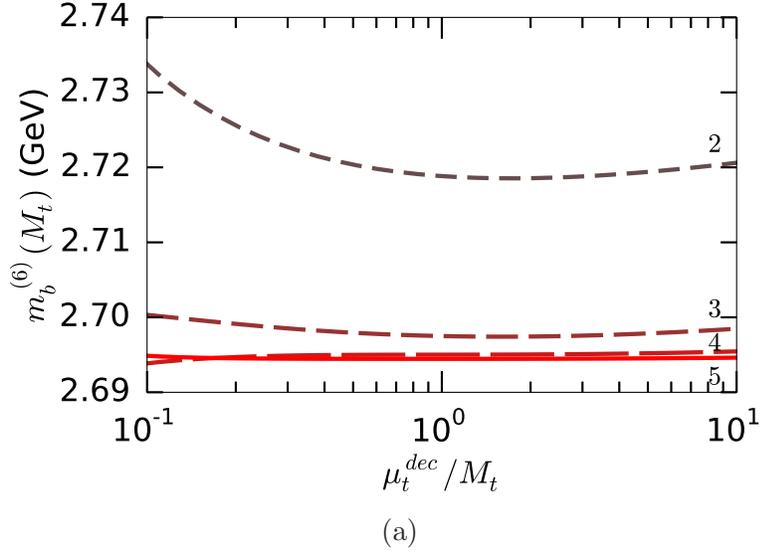}
    \caption[]{\label{fig::mb6mt}}
  \end{subfigure}
  \begin{subfigure}[b]{.7\textwidth}
    \includegraphics[width=\textwidth]{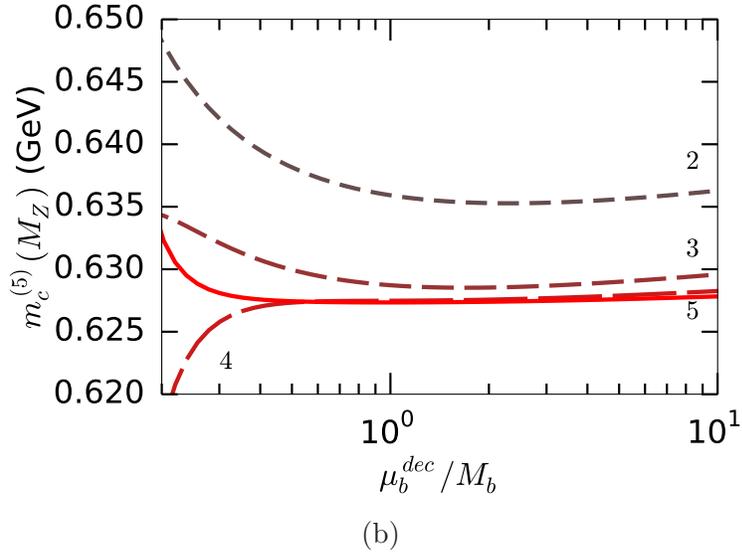}
    \caption[]{\label{fig::mc5mz}}
  \end{subfigure}
  \caption{$m_b^{(6)}(M_t)$ as a function of $\mu_t^{\rm dec}$ (a) and
    $m_c^{(5)}(M_Z)$ as a function of $\mu_b^{\rm dec}$ (b).  The numbers
    indicate the loop order used for the running.}
\end{figure}

As a first phenomenological application we consider the evaluation of the
bottom quark mass at the scale of the top quark with six active flavours using
$m_b^{(5)}(m_b^{(5)})$ as input. We are interested in the dependence of
$m_b^{(6)}(M_t)$ on the decoupling scale of the top quark. Since this scale is
unphysical it should get weaker after including higher order corrections.  Our
results, which are shown in Fig.~\ref{fig::mb6mt}, are obtained using the
following scheme, where $N\in\{1,2,3,4,5\}$ refers to the number of loops:
\begin{itemize}
\item Use $N$-loop running:
  $m_b^{(5)}(m_b^{(5)}) \to m_b^{(5)}(\mu_t^{\rm dec})$
\item Use $(N-1)$-loop decoupling:
  $m_b^{(5)}(\mu_t^{\rm dec}) \to m_b^{(6)}(\mu_t^{\rm dec})$
\item Use $N$-loop running
  $m_b^{(6)}(\mu_t^{\rm dec}) \to m_b^{(6)}(M_t)$
\end{itemize}
The values for $\alpha_s$ involved in this procedure,
$\alpha_s^{(5)}(m_b^{(5)}(m_b^{(5)}))$,
$\alpha_s^{(5)}(\mu_t^{\rm dec})$,
$\alpha_s^{(6)}(\mu_t^{\rm dec})$, and
$\alpha_s^{(6)}(M_t)$,
are obtained from $\alpha_s^{(5)}(M_Z)$ using the same 
loop-order for the running and decoupling as described above
for the bottom quark mass.

In Fig.~\ref{fig::mb6mt} $m_b^{(6)}(M_t)$ is shown as a function of the scale
$\mu_t^{\rm dec}$ where the transition from five- to six-flavour QCD is
performed normalized to the on-shell top quark mass.  
For the 
on-shell top quark mass we choose $M_t=173.34$~GeV~\cite{ATLAS:2014wva}.
We vary $\mu_t^{\rm
  dec}/M_t$ by a factor of 10 around the central scale $\mu_t^{\rm
  dec}/M_t=1$. The one-loop result leads to $m_b^{(6)}(M_t)\approx 2.9$~GeV
and is not shown in the plot. One observes that already the result where
two-loop running is used (short-dashed line) shows only a weak dependence on
$\mu_t^{\rm dec}$. It becomes even weaker at three and four loops (results
with higher perturbative order have longer dashes) and results in an almost
flat curve at five loops (solid line) which can barely be distinguished from
the four-loop curve.  The five-loop results depends on the unknown five-loop
coefficient $\beta_4$ of the beta function.  Our
default choice in Fig.~\ref{fig::mb6mt} is $\beta_4=100\beta_0$ ($\beta_0=
11/4 - n_f/6$) which is numerically close to the Pad\'e estimate obtained
  in Ref.~\cite{Ellis:1997sb}.
For $\beta_4=0$ and $\beta_4=200\beta_0$ one observes a shift
of the five-loop result by about $+0.5$~MeV and $-0.5$~MeV, respectively.

It is interesting to look at the shift on $m_b^{(6)}(M_t)$ at the central
scale $\mu_{\rm dec} = M_t$. The two-, three- and four-loop curves 
lead to shifts of about $-201$~MeV, $-21$~MeV and $-2$~MeV, respectively.
For $\beta_4 = 100\beta_0$ the five-loop result leads to a shift
of about $-0.5$~MeV.

In a second application we consider the evaluation of $m_c^{(5)}(M_Z)$ with
$m_c^{(4)}(3~\mbox{GeV})$ as input.  The calculation proceeds in analogy to
the bottom quark case discussed before, where for the on-shell bottom quark
mass we use the value $M_b=4.7$~GeV. Our results are shown in
Fig.~\ref{fig::mc5mz}.  Again one observes a flattening of the curves after
including higher order corrections. However, for $\mu_b^{\rm dec} \approx
1$~GeV, which corresponds to the left border of Fig.~\ref{fig::mc5mz}, all
curves show a strong variation which indicates the breakdown of perturbation
theory for small scales. Around $\mu_b^{\rm dec}/M_b \gsim 0.3$ both the
four- and five-loop curves are basically flat.

At the central scale $\mu_b^{\rm dec} = M_b$ one observes shifts in 
$m_c^{(5)}(M_Z)$ of $-55$~MeV, $-7$~MeV and $-1$~MeV after including two-,
three- and four-loop running accompanied by one-, two- and three-loop
decoupling. The shift at five loops is below 1~MeV
for $\beta_4 = 100\beta_0$ but also for
$\beta_4 = 0$ and $\beta_4 = 200\beta_0$.


\section{\label{sec::let}Low-energy theorem: Higgs-fermion coupling}

The effective Lagrangian describing the coupling of a Higgs boson 
to gluons and light quarks can be written in the form
\begin{eqnarray}
  {\cal L}_{\rm eff}&=&-\frac{H^0}{v^0}
  \left(C_1  {\cal O}_1^\prime + C_2 {\cal O}_2^\prime \right)
  \,,
  \label{eq::Leff}
\end{eqnarray}
where the effective operators, which are
constructed from light degrees of freedom~\cite{Spiridonov:1984br},
are given by
\begin{eqnarray}
  {\cal O}^\prime_1&=& \left(G^{a,\mu\nu}\right)^2\,,
  \nonumber\\
  {\cal O}^\prime_2&=&
  \sum_{i=1}^{n_l}m_{q_i}^{0\prime}\bar\psi_{q_i}^{0\prime}\psi_{q_i}^{0\prime}
  \,.
  \label{eq::O2}
\end{eqnarray}
The residual dependence on the mass $m_h$ of the heavy quark $h$ is contained
in the coefficient functions $C_1^0$ and $C_2^0$.  In Eq.~(\ref{eq::Leff}) $H$
denotes the Higgs field and $v$ the vacuum expectation value. The superscript
``0'' reminds us that the corresponding quantities are bare.  For the
renormalization of $C_1^0, C_2^0, {\cal O}_1^\prime$ and ${\cal O}_2^\prime$
we refer to Ref.~\cite{Spiridonov:1984br,Chetyrkin:1997un}; for the purpose of
this paper it is of no further relevance. In Ref.~\cite{Chetyrkin:1997un} a
low-energy theorem has been derived which relates the computation of the
renormalized coefficient function $C_2$ to derivatives of $\zeta_m$ w.r.t. the
heavy mass $m_h$. It is given by
\begin{eqnarray}
  C_2 &=& 1 + \frac{\partial\ln\zeta_m}{\partial \ln m_h}\,.
  \label{eq::c2let}
\end{eqnarray}
It should be stressed that Eq.~(\ref{eq::c2let}) is valid to all orders in
$\alpha_s$.  Note that Eq.~(\ref{eq::c2let}) contains the derivative w.r.t.
$\ln m_h$ and furthermore the $m_h$ dependence of $C_2$ appears in the form
$\ln(\mu/m_h)$. Thus we can exploit renormalization group techniques to
construct all logarithmic terms of the next, not computed perturbative
order. In particular, on the basis of our four-loop calculation for $\zeta_m$
we can compute $C_2$ to five-loop accuracy using the recently computed
five-loop result for the quark mass anomalous dimension~\cite{Baikov:2014qja}.
Note that the four-loop anomalous dimensions have been computed in
Refs.~\cite{Chetyrkin:1997dh,Vermaseren:1997fq} ($\gamma_m$) and
Refs.~\cite{vanRitbergen:1997va,Czakon:2004bu} ($\beta$), respectively.

Inserting $\zeta_m^{\overline{\rm MS}}$ into Eq.~(\ref{eq::c2let})
we obtain the following result
\begin{eqnarray}
  C_2^{\overline{\rm MS}} &=& 1 
  + \api^2 0.2778 
  + \api^3 \( 2.243 + 0.2454 \,n_l\) 
  \nonumber\\&&\mbox{}
  + \api^4 \(2.180 + 0.3096 \,n_l - 0.01003 \,n_l^2\) 
  \nonumber\\&&\mbox{}
  + \api^5 \(66.71 + 13.44 \,n_l - 3.642 \,n_l^2 + 
  0.07556 \,n_l^3\)
  \,,
\end{eqnarray}
where we have chosen $\mu=m_h$ to obtain more compact expressions.
Analytic result valid for general $\mu$ are provided
from~\cite{progdata}.

In practice, one often encounters the situation where $C_2$ has to be inserted
in a formula expressed in terms of $\alpha_s^{(n_l)}$. If we furthermore
transform the heavy quark mass to the on-shell scheme
we obtain for $\mu=M_h$
\begin{eqnarray}
  C_2^{\rm OS} &=& 1 
  + \apinl^2 0.2778 
  + \apinl^3 \(1.355 + 0.2454 \,n_l\)
  \nonumber\\&&\mbox{}
  + \apinl^4 \(-12.13 + 1.004 \,n_l - 0.01003 \,n_l^2\)
  \nonumber\\&&\mbox{}
  + \apinl^5 \(-140.9 + 44.20 \,n_l - 4.332 \,n_l^2 + 0.07556 \,n_l^3\)
  \,.
  \label{eq::C2OS5}
\end{eqnarray}

Let us briefly discuss the influence of $C_2$ on the Higgs boson decay to
bottom quarks where the role of the heavy quark is taken over by the top
  quark. We consider the contributions proportional to $(C_2)^2$ from
Eq.~(\ref{eq::Leff}) and use the result for the massless correlator from
Ref.~\cite{Baikov:2005rw}. For convenience we identify the
renormalization scale with the Higgs boson mass and set $\mu=M_H$.  Then
the decay rate of the Standard Model Higgs boson to bottom quarks can be
written in the form
\begin{eqnarray}
  \Gamma(H\to b\bar{b}) &=&
  \frac{G_F M_H^2}{4\sqrt{2}\pi}m_b^2(M_H) \, R(M_H)\,,
  \\
  R(M_H)&=&
  1
  + 5.667 \left(\frac{\alpha_s}{\pi}\right) 
  + \left(29.147 + 0.991\right) 
  \left(\frac{\alpha_s}{\pi}\right)^2 
  +\left(41.758+13.105\right) \left(\frac{\alpha_s}{\pi}\right)^3 
  \nonumber \\&&\mbox{}
  +\left(- 825.7+ 50.7\right) \left(\frac{\alpha_s}{\pi}\right)^4
  + \left(r_5 +  224.8\right) \left(\frac{\alpha_s}{\pi}\right)^5
  \label{eq::hbb}\\
  &=& 
  1 
  + 0.20400
  + \left(0.03777 + 0.00128\right)
  + \left(0.00195 + 0.00061\right)
  \nonumber\\ &&\mbox{}
  + \left(- 0.00139 + 0.00009\right)
  + \left(0.00000006 r_5 + 0.00001\right)\,, 
  \nonumber
\end{eqnarray}
with $\alpha_s\equiv\alpha_s(M_H)\approx0.1131$.  The first number in the
round brackets in Eq.~(\ref{eq::hbb}) corresponds to the case
$C_2=1$~\cite{Baikov:2005rw} and the second one to the contribution from
$(C_2-1)$.  At three-loop order the top quark induced part amounts to about
30\%, at order $\alpha_s^4$ only 6\%.  Note that the massless correlator at
order $\alpha_s^5$, denoted by $r_5$ in Eq.~(\ref{eq::hbb}),
is currently unknown. 
The $\alpha_s^5$ term in Eq.~(\ref{eq::hbb}) origins from the
five-loop contribution in Eq.~(\ref{eq::C2OS5}) and products of lower-order
contributions. 

Note that in this consideration the contribution of $C_1$
(cf. Eq.~(\ref{eq::Leff})) has been neglected. The corresponding corrections
of order $\alpha_s^3$ can be found in
Ref.~\cite{Chetyrkin:1997vj}. Corrections of order $\alpha_s^4$ which are
proportional to $C_1C_2$ require the evaluation of massless four-loop
two-point functions and are currently unknown.  
Corrections of order $\alpha_s^5$ to the Higgs boson decay rate
involving $(C_1)^2$ have been computed in Ref.~\cite{Baikov:2006ch}.

In Refs.~\cite{Schroder:2005hy,Chetyrkin:2005ia} the five-loop result for
$C_1$ is given in terms of $\alpha_s^{(n_f)}$ and the $\overline{\rm MS}$
quark mass.  We complement this result by $C_1$ parametrized in terms of 
the effective coupling constant and the on-shell mass:
\begin{eqnarray}
  C_1^{\rm OS} &=& -\frac{1}{12} \frac{\alpha_s^{(n_l)}}{\pi} \left\{
    1 + \apinl  2.750 
    + \apinl^2 \(9.642 - 0.6979 n_l\) 
    \nonumber\right.\\&&\left.\mbox{}
    + \apinl^3 \(50.54 - 6.801 n_l - 0.2207 n_l^2\) 
    + \apinl^4 \bigg[-625.2 + 149.8 n_l 
      \nonumber\right.\\&&\left.\mbox{}
      - 3.090 n_l^2 - 0.07752 n_l^3
      + 6 \( \beta_4^{(n_l)} - \beta_4^{(n_l+1)} \) \bigg]
  \right\}\,,
\end{eqnarray}
where $\mu=M_t$ has been chosen.
The analytic version in computer-readable form can again be found
in~\cite{progdata}.


\section{\label{sec::con}Summary and conclusions}

In this paper we compute the four-loop corrections to the decoupling constant
for light quark masses, $\zeta_m$, which has to be applied every time heavy
quark thresholds are crossed. It constitutes a fundamental constant of QCD and
accompanies the five-loop quark anomalous dimension~\cite{Baikov:2014qja} in
the ``running and decoupling'' procedure.  Our results complete the
calculation of the four-loop decoupling constants which has been started in
Refs.~\cite{Schroder:2005hy,Chetyrkin:2005ia}. Note that the five-loop
corrections to the QCD beta function, which is needed to establish relations
between $\alpha_s(\mu)$ and $m_q(\mu)$ at low and high energy scales, is still
missing.

As a by-product of our calculation we obtain the effective coupling of a
scalar Higgs boson and light quarks to five-loop order. It is obtained from
$\zeta_m$ with the help of an all-order low-energy theorem.
We briefly investigate the influence on $\Gamma(H\to b\bar{b})$.



\section*{Acknowledgements}

We would like to thank Konstantin Chetyrkin for useful comments to the
manuscript. We are grateful to Roman Lee for providing us with 
an analytic expression for the $\epsilon^3$ term of $J_{6,2}$
as defined in Ref.~\cite{Lee:2010hs}.



\begin{thebibliography}{99}


%
%

\bibitem{Baikov:2015tea}
  P.~A.~Baikov, K.~G.~Chetyrkin and J.~H.~K\"uhn,
  arXiv:1501.06739 [hep-ph].

\bibitem{Chetyrkin:2015mxa}
  K.~G.~Chetyrkin, J.~H.~K\"uhn, M.~Steinhauser and C.~Sturm,
  arXiv:1502.00509 [hep-ph].

\bibitem{Chetyrkin:1997dh}
  K.~G.~Chetyrkin,
  Phys.\ Lett.\ B {\bf 404} (1997) 161
  [hep-ph/9703278].

\bibitem{Vermaseren:1997fq}
  J.~A.~M.~Vermaseren, S.~A.~Larin and T.~van Ritbergen,
  Phys.\ Lett.\ B {\bf 405} (1997) 327
  [hep-ph/9703284].

\bibitem{vanRitbergen:1997va}
  T.~van Ritbergen, J.~A.~M.~Vermaseren and S.~A.~Larin,
  Phys.\ Lett.\ B {\bf 400} (1997) 379
  [hep-ph/9701390].

\bibitem{Czakon:2004bu}
  M.~Czakon,
  Nucl.\ Phys.\  B {\bf 710} (2005) 485
  [hep-ph/0411261].

\bibitem{Chetyrkin:2004mf}
  K.~G.~Chetyrkin,
  Nucl.\ Phys.\ B {\bf 710} (2005) 499
  [hep-ph/0405193].

\bibitem{Baikov:2014qja}
  P.~A.~Baikov, K.~G.~Chetyrkin and J.~H.~K\"uhn,
  JHEP {\bf 1410} (2014) 76
  [arXiv:1402.6611 [hep-ph]].

\bibitem{Schroder:2005hy}
  Y.~Schroder and M.~Steinhauser,
  JHEP {\bf 0601} (2006) 051
  [hep-ph/0512058].

\bibitem{Chetyrkin:2005ia}
  K.~G.~Chetyrkin, J.~H.~K\"uhn and C.~Sturm,
  Nucl.\ Phys.\ B {\bf 744} (2006) 121
  [hep-ph/0512060].

\bibitem{Chetyrkin:1997un}
  K.~G.~Chetyrkin, B.~A.~Kniehl and M.~Steinhauser,
  Nucl.\ Phys.\ B {\bf 510} (1998) 61
  [hep-ph/9708255].

\bibitem{Nogueira:1991ex}
  P.~Nogueira,
  J.\ Comput.\ Phys.\  {\bf 105} (1993) 279.

\bibitem{Vermaseren:2000nd}
  J.~A.~M.~Vermaseren,
  arXiv:math-ph/0010025.

\bibitem{Kuipers:2012rf}
  J.~Kuipers, T.~Ueda, J.~A.~M.~Vermaseren and J.~Vollinga,
  Comput.\ Phys.\ Commun.\  {\bf 184} (2013) 1453
  [arXiv:1203.6543 [cs.SC]].

\bibitem{Harlander:1997zb}
  R.~Harlander, T.~Seidensticker and M.~Steinhauser,
  Phys.\ Lett.\ B {\bf 426} (1998) 125
  [hep-ph/9712228].

\bibitem{Seidensticker:1999bb}
  T.~Seidensticker,
  hep-ph/9905298.

\bibitem{Laporta:2001dd}
  S.~Laporta,
  Int.\ J.\ Mod.\ Phys.\ A {\bf 15} (2000) 5087
  [hep-ph/0102033].

\bibitem{Smirnov:2014hma}
  A.~V.~Smirnov,
  arXiv:1408.2372 [hep-ph].

\bibitem{Lee:2010hs}
  R.~N.~Lee and I.~S.~Terekhov,
  JHEP {\bf 1101} (2011) 068
  [arXiv:1010.6117 [hep-ph]].

\bibitem{Laporta:2002pg}
  S.~Laporta,
  Phys.\ Lett.\ B {\bf 549} (2002) 115
  [hep-ph/0210336].

\bibitem{Schroder:2005va}
  Y.~Schroder and A.~Vuorinen,
  JHEP {\bf 0506} (2005) 051
  [hep-ph/0503209].

\bibitem{Chetyrkin:2006dh}
  K.~G.~Chetyrkin, M.~Faisst, C.~Sturm and M.~Tentyukov,
  Nucl.\ Phys.\ B {\bf 742} (2006) 208
  [hep-ph/0601165].

\bibitem{priv_R.Lee}
  R.~Lee, private communication.

\bibitem{Steinhauser:2000ry}
  M.~Steinhauser,
  Comput.\ Phys.\ Commun.\  {\bf 134} (2001) 335
  [hep-ph/0009029].

\bibitem{Grozin:2007fh}
  A.~G.~Grozin, P.~Marquard, J.~H.~Piclum and M.~Steinhauser,
  Nucl.\ Phys.\ B {\bf 789} (2008) 277
  [arXiv:0707.1388 [hep-ph]].

\bibitem{Grozin:2011nk}
  A.~G.~Grozin, M.~Hoeschele, J.~Hoff, M.~Steinhauser,
  JHEP {\bf 1109} (2011) 066
  [arXiv:1107.5970 [hep-ph]].

\bibitem{Gray:1990yh}
  N.~Gray, D.~J.~Broadhurst, W.~Grafe and K.~Schilcher,
  Z.\ Phys.\ C {\bf 48} (1990) 673.

\bibitem{Chetyrkin:1999qi}
  K.~G.~Chetyrkin and M.~Steinhauser,
  Nucl.\ Phys.\ B {\bf 573} (2000) 617
  [hep-ph/9911434].

\bibitem{Melnikov:2000qh}
  K.~Melnikov and T.~v.~Ritbergen,
  Phys.\ Lett.\ B {\bf 482} (2000) 99
  [hep-ph/9912391].

\bibitem{progdata}
{\tt http://www.ttp.kit.edu/Progdata/ttp15/ttp15-005}

\bibitem{Agashe:2014kda}
  K.~A.~Olive {\it et al.}  [Particle Data Group Collaboration],
  Chin.\ Phys.\ C {\bf 38} (2014) 090001.

\bibitem{Chetyrkin:2009fv}
  K.~G.~Chetyrkin, J.~H.~K\"uhn, A.~Maier, P.~Maierhofer, P.~Marquard,
  M.~Steinhauser and C.~Sturm,
  Phys.\ Rev.\ D {\bf 80} (2009) 074010
  [arXiv:0907.2110 [hep-ph]].

\bibitem{ATLAS:2014wva}
  [ATLAS and CDF and CMS and D0 Collaborations],
  arXiv:1403.4427 [hep-ex].

\bibitem{Ellis:1997sb}
  J.~R.~Ellis, I.~Jack, D.~R.~T.~Jones, M.~Karliner and M.~A.~Samuel,
  Phys.\ Rev.\ D {\bf 57} (1998) 2665
  [hep-ph/9710302].

\bibitem{Spiridonov:1984br}
  V.~P.~Spiridonov, preprint
  IYaI-P-0378, 1984.

\bibitem{Baikov:2005rw}
  P.~A.~Baikov, K.~G.~Chetyrkin and J.~H.~K\"uhn,
  Phys.\ Rev.\ Lett.\  {\bf 96} (2006) 012003
  [hep-ph/0511063].

\bibitem{Chetyrkin:1997vj}
  K.~G.~Chetyrkin and M.~Steinhauser,
  Phys.\ Lett.\ B {\bf 408} (1997) 320
  [hep-ph/9706462].

\bibitem{Baikov:2006ch}
  P.~A.~Baikov and K.~G.~Chetyrkin,
  Phys.\ Rev.\ Lett.\  {\bf 97} (2006) 061803
  [hep-ph/0604194].


\end{thebibliography}
\end{document}